# Computational Analysis of .NET Remoting and Mobile agent in Distributed Enviroment

Vivek Tiwari, Shailendra G., Renu Tiwari and Malam k.

**Abstract**—A mobile agent is a program that is not bound to the system on which it began execution, but rather travels amongst the hosts in the network with its code and current execution state (i.e. Distributed Environment).The implementation of distributed applications can be based on a multiplicity of technologies, e.g. plain sockets, Remote Procedure Call (RPC), Remote Method Invocation (RMI), Java Message Service (JMS), .NET Remoting, or Web Services. These technologies differ widely in complexity, interoperability, standardization, and ease of use.  The Mobile Agent technology is emerging as an alternative to build a smart generation of highly distributed systems. . In this work, we investigate the performance aspect of agent-based technologies for information retrieval. We present a comparative performance evaluation model of Mobile Agents versus .Net remoting  by means of an analytical approach.  A quantitative   measurements  are  performed  to  compare  .Net remoting  and  mobile  agents  using  communication time, code size(agent code ),Data size, number of node as performance parameters in this research work. The results  depict that Mobile Agent paradigm offers a superior performance compared to .Net remoting  paradigm, offers fast computational speed; procure lower invocation cost by making local invocations instead of remote invocations over the network, thereby reducing network bandwidth.

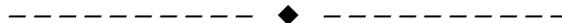

## 1  INTRODUCTION

.Net widely used for creating Middleware solutions. There are many interesting aspects which can be compared between mobile agent and .Net. Mobile computing refers to distributed computing systems that permit user to access various distributed services anywhere and anytime. The diversity of available middleware technologies to implement distributed applications is quite large. These technologies differ widely in the underlying communication paradigm, ease of use for developers, interoperability, and standardization level. Currently, most distributed systems supported for mobile computing are based on classical communication paradigms like RPC (Remote Procedure Call) [4] and Java RMI (Remote Method Invocation) [8] or .NET Remoting [10]. Commonly, a mobile agent is defined  as an independent software program that acts on behalf of a user and helps him to perform tasks. Mobile agents Communication often force the user to disconnect from the network i.e. do tasks on behalf of the user, who meanwhile is totally disconnected.  So my aim is to highlight the benefits  of mobile agents, in particular, their efficiency and flexibility in use. In this work however, we attempt to compare the mobile agent and client server paradigms both in .NET environment using remoting for client server.

The remainder of this paper is organized as follows. Section II presents the related work.  Section III provides a way of working of .NET remoting model.  Section IV presents the working of mobile agent model. Section V present analyses of both .NET remoting and mobile agent with various performance parameters. Section VI show result analysis. Section VII is about implementation of mobile agent using Tahiti server. Section VIII is conclusion.

## 2  RELATED WORKS

Several works have been done in developing mobile agent system for information storage and retrieval, but not much has since been done in the area of comparison of the different approaches with different performance parameter for distributed information management. [13]The author compared the design and the architecture of RMI and .Net remoting  and  analyzed the performance of both technologies using three different scenarios. Elbers et al. [15] have compared the performance of Java RMI and .NET Remoting. In [1] the use of mobile agent in the control and transfer of data in distributed computing environment was presented against the traditional client-server computing using RPC (Remote Procedure Call). The author explores bandwidth utilization, percentage of denial of service, and network overload with retransmission to show a proof of superior scheme provided by mobile agent as against the client server computing. Mobile Agent in Distributed Information Retrieval [3] analyzes base performance in terms of response time (the in-terval between when a request is sent and the time the result is received). The authors made attempt to prove that mobile agent incurs significantly more overhead than RPC due to inter-agent communication and migration when the number of queries are few; however MA begins to perform excellently as the number of queries per query increase.

There have been some initiatives to produce mobile agent systems using the Microsoft .NET Framework, namely MAPNET [7] and EtherYatri.NET. Some systems, such as CARLA [12].

## 3 WORKING OF .NET REMOTING

.Net Remoting is a mechanism which allows the user to invoke methods in another address space. These two mechanisms can be regarded as RPC in an object-oriented fashion. Distributeb object an object (with state and methods), whose interface and implementation reside on different machines.

The .NET Framework is a platform that simplifies application development in the highly distributed environment of the Internet. As opposed to the virtual machine in Java, the .NET version of a *virtual machine* is the CLR. The topic of the CLR (Common Language Runtime) is not within the scope of this paper. Remoting is a framework that is built on the CLR for building distributed applications in an object oriented way. Client applications can invoke functions and access resources on a server object, be they on the same computer, or a remote computer over a network, or another application domain in the same process. Communication between the client and server object is channeled through a proxy object. The proxy allows the client to make calls to the server, using function calls that seem to be local to the client. Fig. 1 shows .Net remoting architecture.

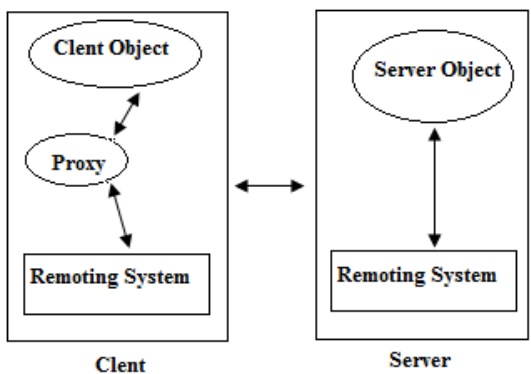

1. Dot Net Remoting Architecture

.Net use distributed objects; these are objects whose instances can be used remotely. In .Net Remoting a Windows native register is used, hence there is no distributed directory within the .Net framework. Proxies in Remoting are created during runtime.

## 4 WORKING OF MOBILE AGENT MODEL

Mobile agent is an object that migrates through many nodes of a heterogeneous network of computers under its own control in order to perform tasks using resources of these nodes [1], [6]. The agents take the itinerary data, execution state and necessary code as part of a "bag" for starting its journey on each server. Once the mobile agent finishes its activities on the server, it determines dynamically the next host to visit. Each node serves as the server for the client to establish communication channel that is based on Transmission Control Protocol/internet Protocol (TCP/IP) sockets [6]. After establishing the connection, the mobile agent queries the local environment to acquire the necessary information to achieve its objectives [2], [9]. The information retrieval service (corresponding to the current node) interacts with its own database management systems and returns the results to the agent, which adds them to its "bag" and migrates to the next host. In case a server is disconnected or faulty, the agent skips the node, determine alternative route dynamically and continues its journey. When all nodes have been visited the agent sends a message to the original host with all results obtained and reports of its itinerary. Figure 2(a) depicts working of .NET remoting model and 2(b) for mobile agent model.

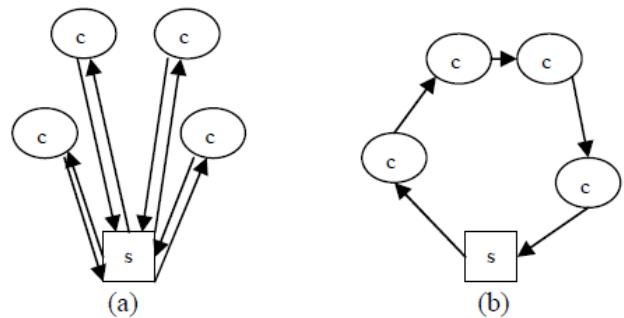

C: client,  S: Server
2: (a) .NET Remoting (b) Mobile Agent

## 5 ANALYSIS OF PERFORMANCE PARAMETERS

Simulation program was developed to provide performance analysis between the .NET remoting and MA. The following parameters were used: Communication *time* and data size cost invocation cost.

### 5.1 Communication Time

The time interval between when a query is sent from the source and when the response (data) is received at the requesting host is referred to as the communication time in this research work. The communication time is measured in terms of request time, **treq** and response time, **tres**. In this paper, the sizes of request and response are assumed to be the same and thus request time and response time are the same and correspond to **tr** [13][15] With .NET remoting when a query is generated, client connects to a remote object that was previously registered on the server, searches through the registered object on the server and reports back to the client. If the required information or object is not found on a host, a report is sent to the client before it connects to another server to search for the same object. It reports back to the server again before connecting to yet another server until the object is located or otherwise [12][10] If we assume the following:

- The distance between all the nodes is equal.
- The time taken to send a query is **treq** and
- The time the response is received at the host is **tres**
- Time required to search distributed object in address space only once.

The search time for one request is the time required to transmit the request from the source node to the destination node and back to the source. This is given by:

$$T^n_{Com} DNR = treq + tres + T_{obj} \quad (1)$$

Where
1. $T_{Com} DNR$ is the communication time for DOT NET remoting (DNR)
2. **treq** is the request time
3. **tres** is the response time
4. $T_{obj}$ time required to search distributed object in address space.

Each time the .NET remoting visits a node it reports back to the requesting host (client), assuming there are n denotes the number of nodes visited during the search, including the launching (client) node, then communication time for **n** node (server) is:

$$T^n_{Com} DNR = n (treq + tres) + T_{obj} \quad (2)$$

Assuming size of request is equal to the size of response and equal to tr. Because .Net Remoting use Windows native register in local so this time is negligible. There is no distributed directory within the .Net framework.

$$T^n_{Com} DNR = 2ntr \quad (3)$$

With mobile agent on the other hand, the client creates an agent which contains the client's request to be executed [12],[14].This agent moves to server A for local interaction, if the object is not found, the agent moves to the next server, B and yet another server until the object is found. Otherwise, until all the servers are visited, before reporting to the client [5],[8]. The agent visits each node in **tr** time, using the same notations, as with DNR, and let us assume the size of agent code is **y.**

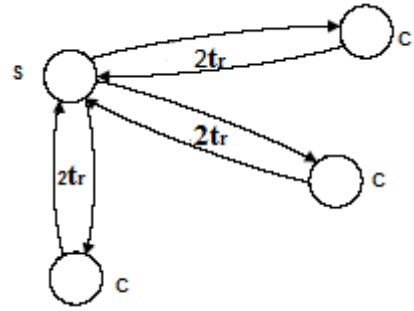

3: Communication Cost in .NET Remoting Model

If it takes **ty** units of time to transfer the agent code on the network. The search time of mobile agent paradigm for one request will be equal to the time to transport the request and agent code on the network between the nodes, thus,

$$T_{Com}AM = tr + ty \quad (4)$$

Assuming the resource is found at node **n,** which includes the launching node, the total search time thus becomes:

$$T^n_{Com}AM = (n-1)(tr + ty) \quad (5)$$

At the end of the search the agent returns to the source with the response in another ( tr + ty ) time. Such that the total time becomes:

$$T^n_{Com}AM = n (tr + ty) \quad (6)$$

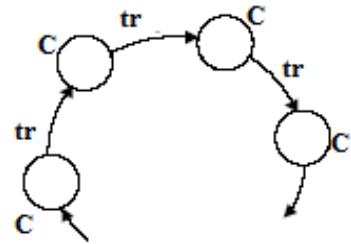

4: Communication Cost in Mobile agent model

### 5.2 Data size cost in Invocation

Cost is estimated in terms of the number and type of invocations involved. The invocations involved in both schemes are classified as either network (remote) invocation or local invocation; we thus generate a cost function for the two schemes. The .NET remoting scheme involves several invocations between the client and the server on the network. In .NET remoting, once a client object connects to a remote object that was registered on a server, the connection has to be maintained to maintain execution[11] .NET remoting makes remote invocations over the network. The request is sent to the server over the network and the response is sent back to the client the same way. The request is taken one after another and sent to the server with the server sending the response each time across the network [7][13] If there are n requests (query) there will be n invocations to the destination and n invocations back to the source. If we assume a request of size ($\alpha$) bytes, response of size ($\beta$) bytes and the cost of one request over the network (network invocation) is ($\psi$), then, for n number of requests, each has a network invocation (n$\psi$) to its destination and one response over the network back to its origin, hence the total costs will be :

$$CDNR = n\alpha + n\beta + n\psi \quad (7)$$

If we assume that the request and response are the same size for the purpose of simplicity, i.e.
($\alpha = \beta$), then the total cost becomes:

$$CDNR = 2n\alpha + n\psi \quad (8)$$

Proxies in .NET Remoting are created during runtime. In our formulas, the proxy download cost is denoted by $T_{proxy}$. As the same remote procedure interface is used in each server, the proxy is downloaded to the client only during the first remote call [8].Now equation 8 became as:

$$CDNR = 2n\alpha + T_{proxy} + n\psi \quad (9)$$

The mobile agent paradigm allows remote invocations to be replaced by local invocations as the agent migrates to the invoked server site [4]. The agent migrates with the execution code and execution thread. The execution thread enables the agent to trace its execution state and execute accordingly. In mobile agent scheme, one network invocation would carry all the requests and mobile agent's code and then execute at the remote server. All other invocations are mere local invocations on the server system. The mobile agent system makes fewer remote invocations cost compared to .NET remoting with a lot of local invocations. For **n** number of requests with size $a'$ bytes, the cost of network invocation will be **n$a'$**. The response to the requesting host is also of size $\beta'$ bytes. If the cost of one local invocation is ($\sigma$) units, assume each request has one local invocation, **n** requests will be **n$\sigma$** units local invocations at the invoked server site. Since the agent moves with its code and execution thread, the size of the mobile agent code will have to be included. Let the size of the mobile agent code be **C** Then the total cost for the mobile agent system will be the sum of the total cost for network invocations, the total local invocations on the invoked server, the mobile code and the response cost over the network.

$$Cma = na' + n\sigma + c + \beta' \quad (10)$$

If the assumption that the size of request and response over the network are equal and equivalent to the cost of network invocations still holds, then, the total cost become:

$$Cma = (n+1)a' + C + n\sigma \quad (11)$$

Where n is the number of requests
$a'$: the size of request over the network
$\beta'$: the size of response over the network
$\sigma$: the cost of local invocation
C: the size of the mobile agent code.

## 6 RESULT ANALYSIS

It is very clear with equation (3) and (6) that communication time taken by .NET remoting is greater than communication time of mobile agent for large n because here **ty** is consider as constant which is smaller than **tr**.

**Since** $\quad t_y < t_r$
$\quad (t_r + t_r) > (t_r + t_y)$
**So**
$$2nt_r > n(t_r + t_y)$$

$$T^n_{Com} DNR > T^n_{Com} AM$$

Now consider equation (9) and (11).Let us consider n is large than data size cost in invocation for .NET remoting is more than for mobile agent.
$$n\psi > n\sigma$$
Because $\psi$ is remote invocation and $\sigma$ is local invocation if we consider α and α' is equal than

$$2n\alpha > (n+1)\alpha'$$

Mobile code size is larger than .NET remoting proxy down load $T_{proxy}$ but these two parameter are much less than other parameter ($\alpha, \psi, \sigma$). So result is

$$CDNR > Cma$$

| CHARACTERISTICS | MOBILE AGENT | .NET REMOTING |
|---|---|---|
| Code Size | Large Code size | Relative Small Code Size |
| Server < 5 | Avg. | Best |
| Server > 7 | Best | Avg. |
| Data Size < 100 MB | Avg. | Best |
| Data Size > 110 MB | Best | Avg. |

5: Performance with parameters

Fig. 5 show performance of .NET remoting and mobile agent on different values of parameter. If number of server is more than 4 then mobile agent work well. Similarly if data size is large (more than 110mb) then mobile agent work better.

Fig. 6 and 7 present performance of mobile agent and .NET remoting with different performance parameters which are code size, no of server and data size. Fig. 8 represents performance with parameter communication time versus no of server. Fig. 9 represents that as performance of mobile agent is increase with data size.

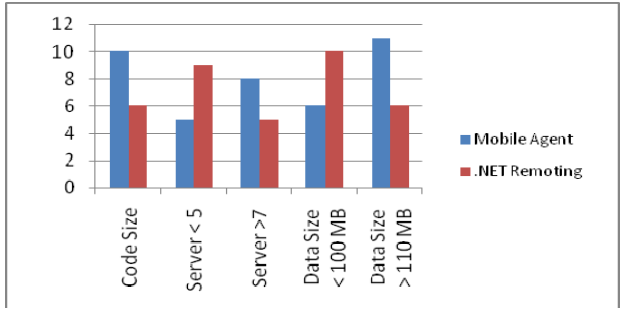

6: Graph 1

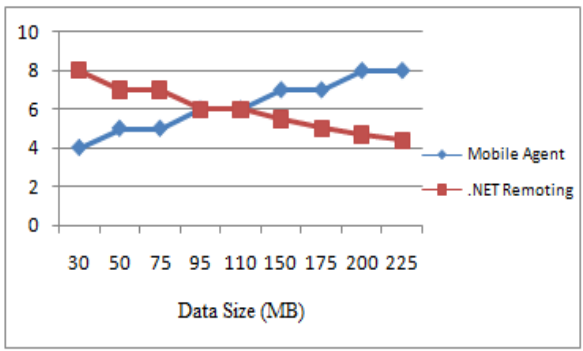

7: Graph 2

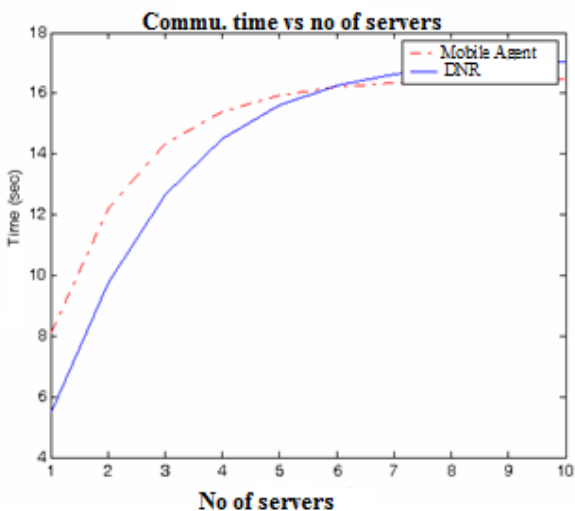

8: Graph 3

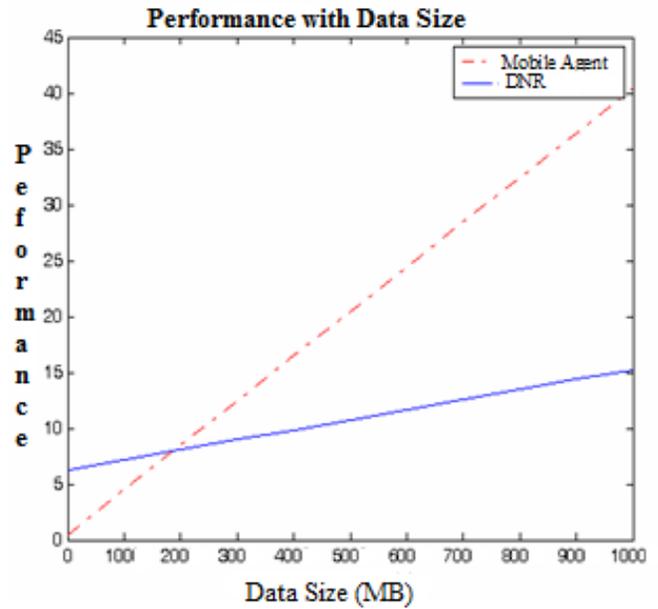

9: Graph 4

## 7 IMPLEMENTATION IN .NET REMOTING

The implementation of .NET Remoting is done on the .NET Framework technology. Whole application is implemented in three modules. First is Server or remotable type second is Listening or host application, and third Client or calling application domain. The server is implemented in RemoteExecuteServer class which contain contains methods to simply run a process. We have created a Windows Forms application as a host application. The host application only needs to configure and register the remote server type. This is done using separate configuration file in XML format. This configuration file should be saved in the same directory as the host application exe file. For a client program: the client must know what methods are available and how to access them. The client program must be compiled with a proxy class to specify the format of the remote class methods. All references to the remote class methods are passed to the proxy class for processing.

## 8 CONCLUSION

The work presented in this paper aims to investigate the use of mobile agents in service provision for mobile environments. The mobile agent applicability for software downloading is assessed by several performance measurements which compare .NET remoting and MA models. The results of this evaluation show that mobile agents provide a significant performance improvement, especially when the number of servers to visit increases or when the amount of data to download is large. Based on these considerable results and in order to illustrate the possibility of using mobile agents for service provision, two enhanced agent-based services, namely service download and service discovery were proposed. Our main perspective is the development of a service provision platform for mobile computing environments based exclusively on mobile agent technology. In this paper, we investigated the use of this technology only for software downloading which represents one of the main requirements of such a platform. Agent-based developments of the remaining requirements such service customization, service provision while roaming, user and terminal mobility, and service adaptability and reconfigurability constitute and network bandwidth and network congestion are a part of our future work.

**Vivek Tiwari.** Asst. Prof. in Faculty of Engineering And Technology, MITS Deemed university, sikar Rajastahan, India. M.Tech. in Computer Science an Engg. from SATI ,Vidisha, India. He did Bechalore of Engineering in Computer Science and Engineering in 2004 from ITM, Gwalior, India.

**Shailendra G.** Asst Prof. in Bhabha Engg College, Bhopal, India. M.Tech. in Software system from SATI ,Vidisha, India. He did Bechalore of Engineering in Computer Science and Engineering in 2004 from ITM, Gwalior, India.

**Renu Tiwari** M.Tech. (final year Student) in Software Engineering, LNCT, Bhopal, India. She did Master of computer application in 2007 from MPCT, Gwalior, India.

**Malam kirar.** PGT in KV no 2, Bhopal, India**.** M.Tech. in Software system from SATI ,Vidisha, India. He did Master of computer application from Bhopal, India.